\documentclass[letter,traditabstract,longauth]{aa} 

\usepackage{graphicx}
\usepackage{natbib}
\usepackage{txfonts}

\begin{document}
\newcommand{\Zsolar}{\mbox{$\,\rm Z_{\odot}$}}
\newcommand{\Msolar}{\mbox{$\,\rm M_{\odot}$}}
\newcommand{\Lsolar}{\mbox{$\,\rm L_{\odot}$}}
\newcommand{\xs}{$\chi^{2}$}
\newcommand{\dxs}{$\Delta\chi^{2}$}
\newcommand{\xsn}{$\chi^{2}_{\nu}$}
\newcommand{\ls}{{\tiny \( \stackrel{<}{\sim}\)}}
\newcommand{\gs}{{\tiny \( \stackrel{>}{\sim}\)}}
\newcommand{\asec}{$^{\prime\prime}$}
\newcommand{\amin}{$^{\prime}$}
\newcommand{\mstar}{\mbox{$M_{*}$}}
\newcommand{\hi}{H{\sc i}\ }
\newcommand{\hii}{H{\sc ii}\ }
\newcommand{\kms}{$\rm km~s^{-1}$}

\author{
J. I. Davies\inst{7}
\and
M. Baes\inst{1} 
\and
G. J. Bendo\inst{2}
\and
S. Bianchi\inst{3}
\and
D. J. Bomans\inst{4}
\and
A. Boselli\inst{5}
\and
M. Clemens\inst{6}
\and
E. Corbelli\inst{3}
\and
L. Cortese\inst{7}
\and
A. Dariush\inst{7}
\and
I. De Looze\inst{1}
\and
S. di Serego Alighieri\inst{3}
\and
D. Fadda\inst{8}
\and
J. Fritz \inst{1}
\and
D. A. Garcia-Appadoo\inst{9}
\and
G. Gavazzi\inst{10}
\and
C. Giovanardi\inst{3}
\and
M. Grossi\inst{11}
\and
T. M. Hughes\inst{7}
\and
L. K. Hunt\inst{3}
\and
A. P. Jones\inst{12}
\and
S. Madden\inst{13}
\and
D. Pierini\inst{14}
\and
M. Pohlen\inst{7}
\and
S. Sabatini\inst{15}
\and
M. W. L. Smith\inst{7}
\and
J. Verstappen\inst{1}
\and
C. Vlahakis\inst{16} 
\and
E. M. Xilouris\inst{17}
\and
S. Zibetti\inst{18}
}

\institute{
Sterrenkundig Observatorium, Universiteit Gent, KrijgslAAn 281 S9, B-9000 Gent,
Belgium 
\and
Astrophysics Group, Imperial College London, Blackett Laboratory, Prince Consort
Road, London SW7 2AZ, UK 
\and
INAF-Osservatorio Astrofisico di Arcetri, Largo Enrico Fermi 5, 50125 Firenze, Italy 
\and
Astronomical Institute, Ruhr-University Bochum, Universitaetsstr. 150, 44780 Bochum,
Germany 
\and
Laboratoire d'Astrophysique de Marseille, UMR 6110 CNRS, 38 rue F. Joliot-Curie,
F-13388 Marseille, France 
\and
INAF-Osservatorio Astronomico di Padova, Vicolo dell'Osservatorio 5, 35122 Padova,
Italy
\and
Department of Physics and Astronomy, Cardiff University, The Parade, Cardiff, CF24
3AA, UK
\and
NASA Herschel Science Center, California Institute of Technology, MS 100-22,
Pasadena, CA 91125, USA 
\and
ESO, Alonso de Cordova 3107, Vitacura, Santiago, Chile 
\and
Universita' di Milano-Bicocca, piazza della Scienza 3, 20100, Milano, Italy 
\and
Centro de Astronomia e Astrof\'isica da Universidade de Lisboa, Observat\'orio
Astron\'omico de Lisboa, Tapada da Ajuda, 1349-018, Lisboa, Portugal and CAAUL, Observat\'orio Astron\'omico de Lisboa, Universidade de Lisboa,
Tapada da Ajuda, 1349-018, Lisboa, Portugal
\and
Institut d'Astrophysique Spatiale (IAS), B\^atiment 121, Universit\'e Paris-Sud 11 and
CNRS, F-91405 Orsay, France 
\and
Laboratoire AIM, CEA/DSM- CNRS - Universit\'e Paris Diderot, Irfu/Service
d'Astrophysique, 91191 Gif sur Yvette, France 
\and
Max-Planck-Institut fuer extraterrestrische Physik, Giessenbachstrasse, Postfach
1312, D-85741, Garching, Germany
\and
INAF-Istituto di Astrofisica Spaziale e Fisica Cosmica, via Fosso del Cavaliere 100,
I-00133, Roma, Italy 
\and
Leiden Observatory, Leiden University, P.O. Box 9513, NL-2300 RA Leiden, The
Netherlands 
\and
Institute of Astronomy and Astrophysics, National Observatory of Athens, I. Metaxa
and Vas. Pavlou, P. Penteli, GR-15236 Athens, Greece 
\and
Max-Planck-Institut fuer Astronomie, Koenigstuhl 17, D-69117 Heidelberg,  Germany 
}

\title{The Herschel Virgo Cluster Survey: I. Luminosity functions \thanks{Herschel is an ESA space observatory with science instruments provided 
by European-led Principal Investigator consortia and with important participation from NASA.}}

   \date{Submitted to AA Herschel Special Issue}

\abstract{
We describe the Herschel Virgo Cluster Survey (HeViCS) and the first data obtained as part of the Science Demonstration Phase (SDP). The data cover a central 4$\times$4 sq deg region of the cluster. We use SPIRE and PACS photometry data to produce 100, 160, 250, 350 and 500 $\mu$m luminosityfFunctions (LFs) for optically bright galaxies that are selected at 500 $\mu$m and detected in all bands. We compare these LFs with those previously derived using IRAS, BLAST and Herschel-ATLAS data. The Virgo Cluster LFs do not have the large numbers of faint galaxies or examples of very luminous galaxies seen previously in surveys covering less dense environments.}

   \keywords{Galaxies: evolution - Galaxies: clusters - ISM: dust}

	\authorrunning{Davies et al.,}	
   \maketitle

\section{Introduction} 
The Virgo Cluster provides us with a unique opportunity to study in detail a large number of galaxies in the cluster environment. Virgo is probably the most studied cluster of galaxies because of its proximity to the Milky Way - it lies at a distance of $\sim$17 Mpc (Gavazzi et al., 1999 and references therein) with a mean velocity of $\sim$1094 km s$^{-1}$ (Binggeli et al., 1987). It is an Abell richness Class I cluster containing $\sim$2000 optically catalogued galaxies (Virgo Cluster Catalogue (VCC), Binggeli et al., 1985). The Virgo Cluster virial mass is $\approx 2.5 \times 10^{14}$
 M$_{\odot}$ (Girardi et al., 1998; Rines \& Diaferio 2006) as expected for a cluster as rich as this. The cluster contains large amounts of X-ray emitting gas (Boehringer et al., 1994) and shows clear evidence for both sub-structure and non-virialised motions (Gavazzi et al., 1999). From previous observations of galaxy clusters we know that this environment can have a significant effect on the properties of galaxies. The morphological mix is quite different compared to the field and many galaxies may be deficient in atomic gas (Haynes \& Giovanelli, 1984). This deficiency is probably responsible for the lower star-formation rates (Lewis et al., 2002, Gomez et al., 2003), the truncation of star-forming discs (Boselli \& Gavazzi, 2006) and the observed higher metal content of cluster galaxies (Skillman et al., 1996). Being so nearby, the Virgo Cluster has previously been studied in detail at many wavelengths and so there is a wealth of data available, to which we are now able to add the FIR.

This is the first of a series of seven papers, based on HeViCS SDP data, in which we discuss: how the cluster environment truncates the dust discs of spiral galaxies (paper II, Cortese et al., 2010), the dust life-time in early-type galaxies (paper III, Clemens et al., 2010), the spiral galaxy dust surface density and temperature distribution (paper IV, Smith et al., 2010), the properties of metal-poor star-forming dwarf galaxies (paper V, Grossi et al., 2010), the lack of thermal emission from the elliptical galaxy M87 (paper VI, Baes et al., 2010) and the far-infrared (FIR) detection of dwarf elliptical galaxies (paper VII, De Looze et al., 2010). A further paper discusses the spectral energy distributions of HeViCS galaxies together with  others observed as part of the Herschel Reference Survey (Boselli et al., 2010). In this paper we briefly discuss the HeViCS project, describe the data used in the above papers and compare the Virgo FIR LFs with those previously derived.

\begin{figure}
\centering
\includegraphics[scale=0.45]{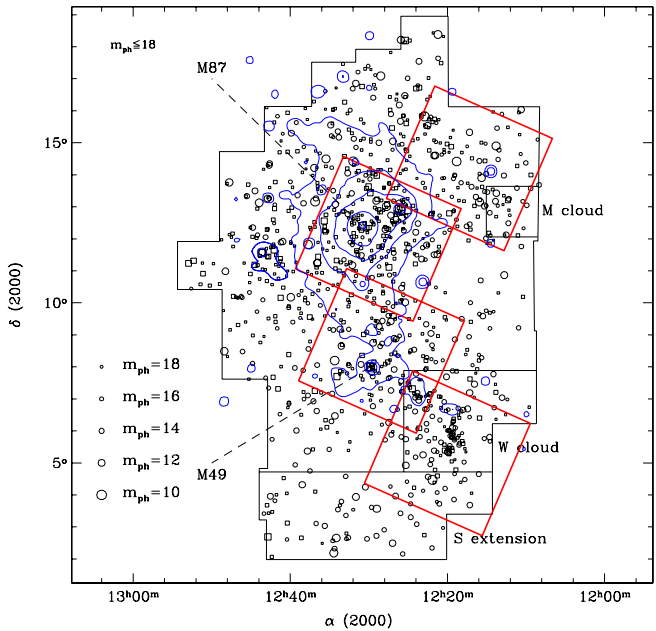}
\caption{
Virgo Cluster. The solid black line indicates the area covered by the VCC, with each VCC galaxy marked by a small black circle. The ROSAT X-ray contours are marked in blue. The area to be observed by HeViCS is indicated by the four red boxes. The SDP data described here are from the field centred on M87.}
\end{figure}

\vspace{-0.2cm}

\section{The HeViCS}
The HeViCS is an approved Open Time Key Project for the
ESA Herschel Space Observatory (Pilbratt et al., 2010). It has
been awarded 286 hours of observing time in parallel mode with
PACS (Poglitsch et al., 2010) at  100 and 160 $\mu$m and SPIRE (Griffin et al., 2010)
at 250, 350 and 500 $\mu$m. We will map four 4$\times$4 sq deg regions of the cluster (Fig. 1) down to the 250 $\mu$m confusion limit of $\approx$ 1 MJy sr$^{-1}$. From within these four regions we expect to detect about 200 galaxies in all bands. At the distance of the Virgo Cluster a typical 50 kpc diameter galaxy subtends an angle of about 10\arcmin. Herschel with resolution of order 10\arcsec is thus able to resolve structure and we can study the detailed FIR properties of these galaxies for the first time. 

The primary HeViCS science goals include the detection of dust in the inter-galactic medium, the extent of cold dust in the outskirts of galaxies, the FIR LFs, the complete spectral energy distributions (SEDs) of galaxies, the dust content of dwarf ellipticals and irregulars and a detailed analysis of the dust content of early type galaxies (for further details see http://www.hevics.org).

\section{The data}
We have obtained SDP data for one field at the centre of the
Virgo Cluster (Fig. 2) using the SPIRE/PACS parallel scan-map mode with nominal
detector settings and fast scan speed (60\arcsec/sec). Two cross-linked
scans were carried out, which is just one quarter of the exposure time of the completed survey.

PACS data reduction was carried with the standard pipeline for both the 100
and 160 $\mu$m channels.
Dead and saturated pixels were masked. Deglitching was performed in two steps, using the standard multi-wavelength median transform 
deglitcher, and one based on
sigma-clipping. 
Bright sources were masked before a high-pass filter was used to reduce $1/f$ noise. 
Finally, the two orthogonal scans
were combined and maps made using the naive map-maker.
The full width half maximum (FWHM) beam sizes are 7.0\arcsec$\times$12.7\arcsec and 11.6\arcsec$\times$15.7\arcsec with pixel sizes of 3.2\arcsec and 6.4\arcsec for the 100
and 160 $\mu$m channels respectively. 

For SPIRE the data were reduced up to Level 1 as
described in Pohlen et al. (2010) except for the deglitching. Instead of
the default wavelet deglitching module we used the newer kappa-sigma
de-glitcher. Rather than the default median baseline subtraction we used a simple
robust linear fit with outlier rejection to all data points in the
timeline. This gave a superior result in terms of residual
striping.
Both scans were then combined to make our final maps using the naive mapper provided.
The FWHM of the SPIRE beams are
18.1\arcsec, 25.2\arcsec, and 36.9\arcsec with pixel sizes of 6\arcsec,
10\arcsec, and 14\arcsec at 250, 350, and 500 $\mu$m, respectively. The uncertainty in the flux calibration is of the order of 15\% (Swinyard et al., 2010). 

The final data used by the seven HeViCS papers have a 1$\sigma$ noise determined from the whole of each image of $\sim$8.6, 4.9, 1.3, 0.6 and 0.3 MJy sr$^{-1}$ at 100, 160, 250, 350 and 500 $\mu$m respectively. 

\begin{figure}
\centering
\includegraphics[width=0.485\textwidth]{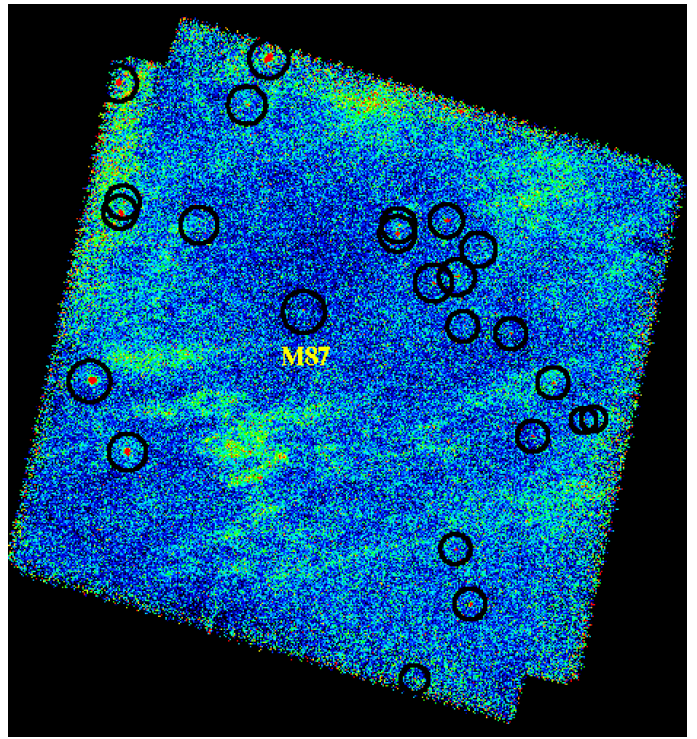}
\caption{
$4^{0}\times4^{0}$ SPIRE 250 $\mu$m SDP data. Circles mark the positions of the galaxies selected with 500 $\mu$m fluxes $>$ 0.2 Jy and an 1.5$\sigma_{500}$ angular diameter $>$ 1.4\arcmin. The cluster centre is marked by the position of M87. Most prominent is the extent of galactic cirrus even at this high Galactic latitude ($b\approx75^{o}$).}
\end{figure}

\section{The luminosity functions}
To obtain LFs we chose to carry out the initial object selection at 500 $\mu$m because it is the least explored part of the spectrum, it has the lowest resolution, and most galaxies will produce their least flux in this band, guaranteeing a detection in all five bands. For the LF analysis the data were smoothed and re-gridded to the 500 $\mu$m resolution and pixel scale. The final 1$\sigma$ noise in the smoothed and re-gridded images are $\sim$2.2, 2.9, 1.2, 0.7 and 0.3 MJy sr$^{-1}$ at 100, 160, 250, 350 and 500 $\mu$m respectively.

To produce an objectively selected sample we used the automatic image detection algorithm SExtractor (Bertin \& Arnouts, 1996). We detected at 500 $\mu$m and then used the same apertures at the other wavelengths.  To minimise background contamination each object had to have more than 30 connected pixels at 1.5$\sigma_{500}$ or above. The result is a 500 $\mu$m flux limit of $\sim$0.2 Jy for sources with diameter larger than 1.4\arcmin and 97 detected objects. Each object was then checked for correspondence with a known Virgo Cluster galaxy. The final list consists of 24 Virgo Cluster objects, just 10\% of the 247 VCC galaxies in the field that are listed by GOLDMINE as definite cluster members (Gavazzi et al., 2003).  
By far the majority of discarded objects appear to be associated with extended galactic cirrus emission. The exception are nine detections that have images that appear to be galaxies but are not in the VCC. In each case they are close to ($\sim10$\arcsec) a faint SDSS galaxy with no redshift, so provisionally we assume these to be objects in the background.

We also visually inspected the positions of all confirmed VCC galaxies in the 500 $\mu$m image to see if any other galaxies could be detected. One hundred and fifty-eight (64\%) of the VCC galaxies in this field are classified dE, one of which is a low signal-to-noise detection at 500 $\mu$m (De Looze et al., 2010). Twenty-eight (11\%) are Sm/Im/BCD/dIrr four of which are detected at 500 $\mu$m, but below our threshold (Grossi et al., 2010). Thirty-seven (15\%) are other spiral and early type galaxies that with the exception of three are again not detected at 500 $\mu$m. These three (1\%) are clearly detected, but fail by their isophotal size. The positions within the cluster of the twenty-four 500 $\mu$m selected galaxies is shown in Fig. 2. 

The LFs at each wavelength are shown in Fig. 3. These FIR LFs are quite different compared to the optical, where large numbers of dwarf galaxies are found at the faint-end. They are similar to the HI mass function, where there is a suggestion of a turnover at low masses (Davies et al., 2004). The comparatively low numbers at the faint end cannot be due to incompleteness because for this bright galaxy sample even the two faintest FIR sources have a signal-to-noise at 500 $\mu$m of $\sim$20. It is clear that there are many low-luminosity galaxies in the cluster that are detected in the optical, but their lack of gas and dust makes them difficult to detect at 21cm and in the FIR. 

\begin{figure}
\centering
\includegraphics[scale=0.38]{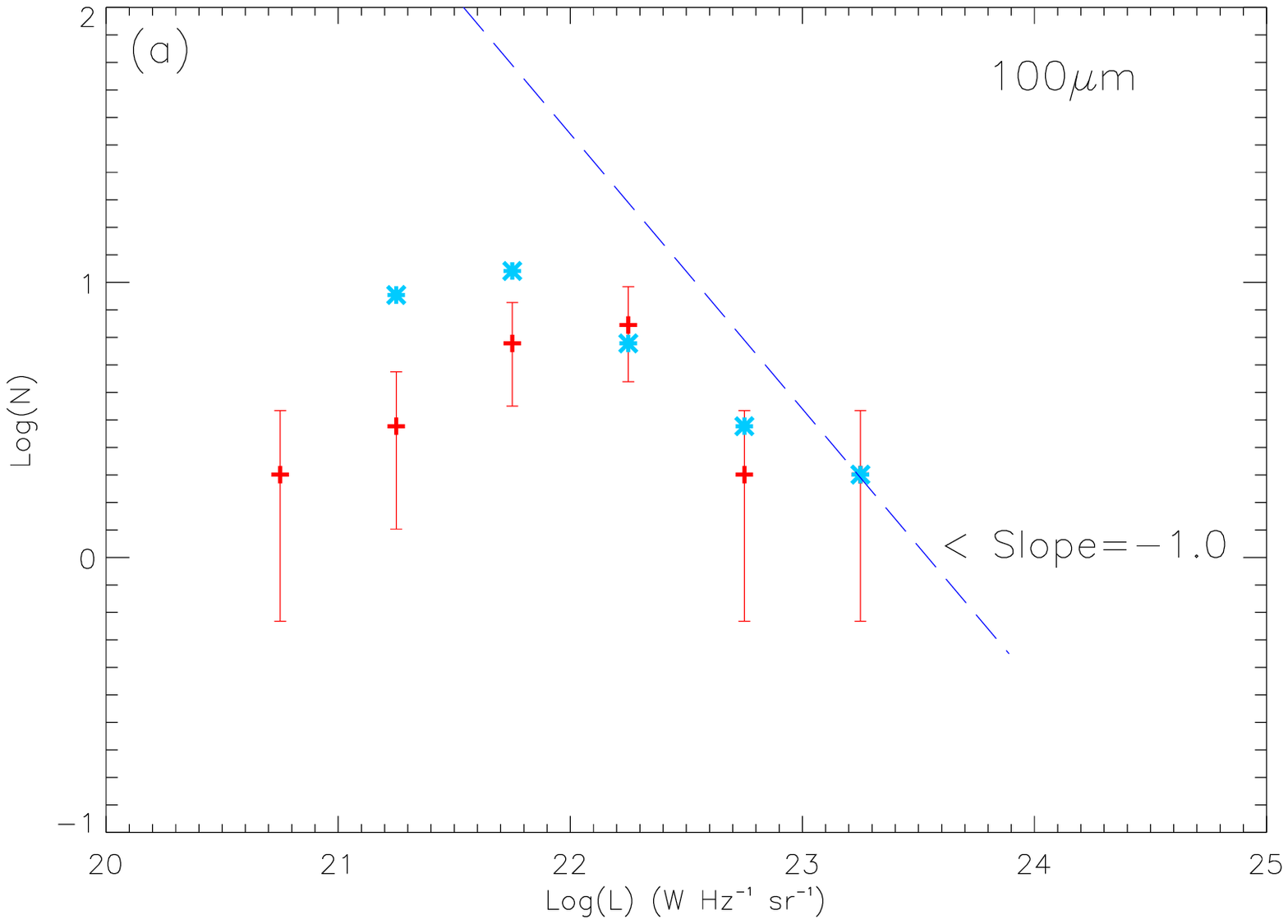}
\includegraphics[scale=0.38]{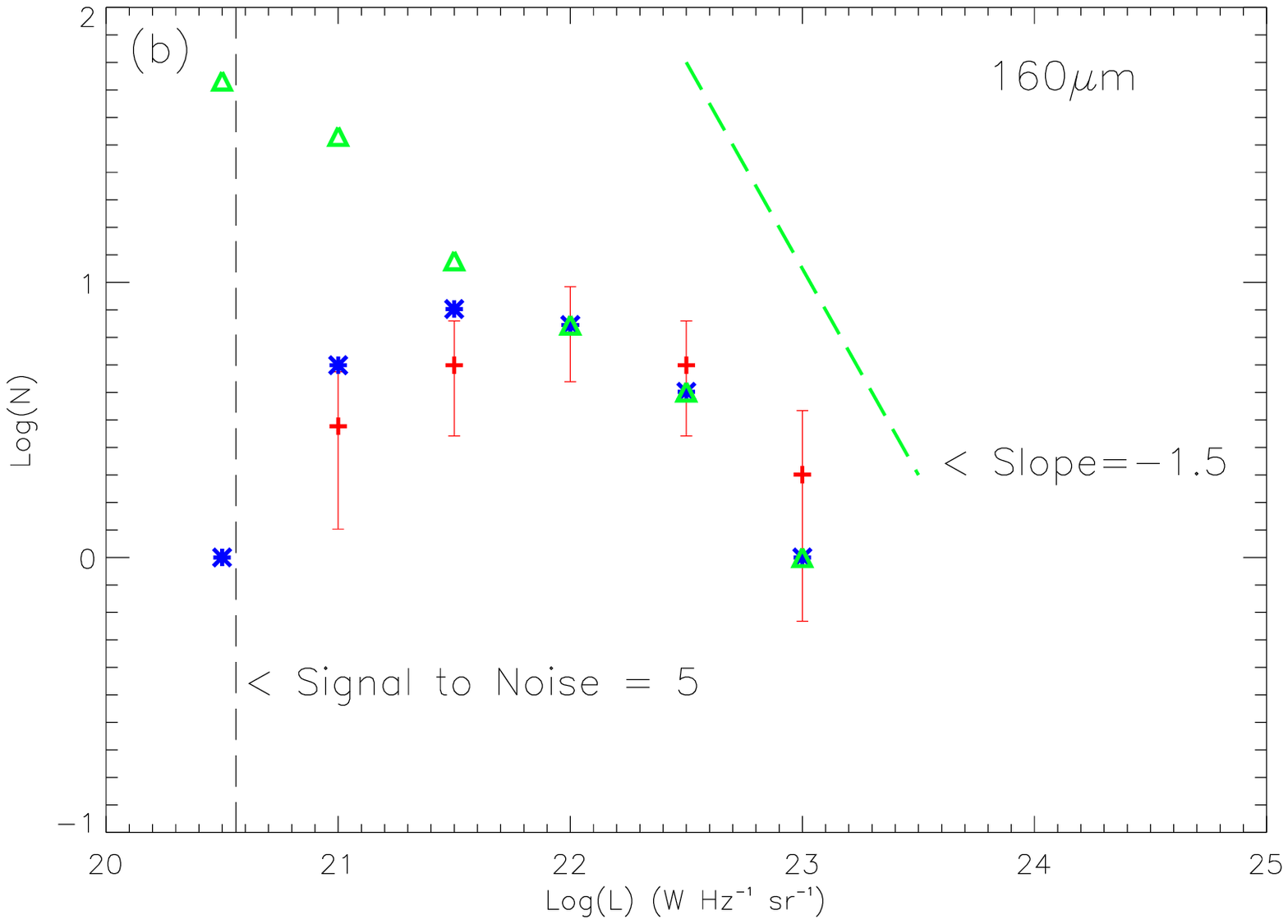}
\includegraphics[scale=0.38]{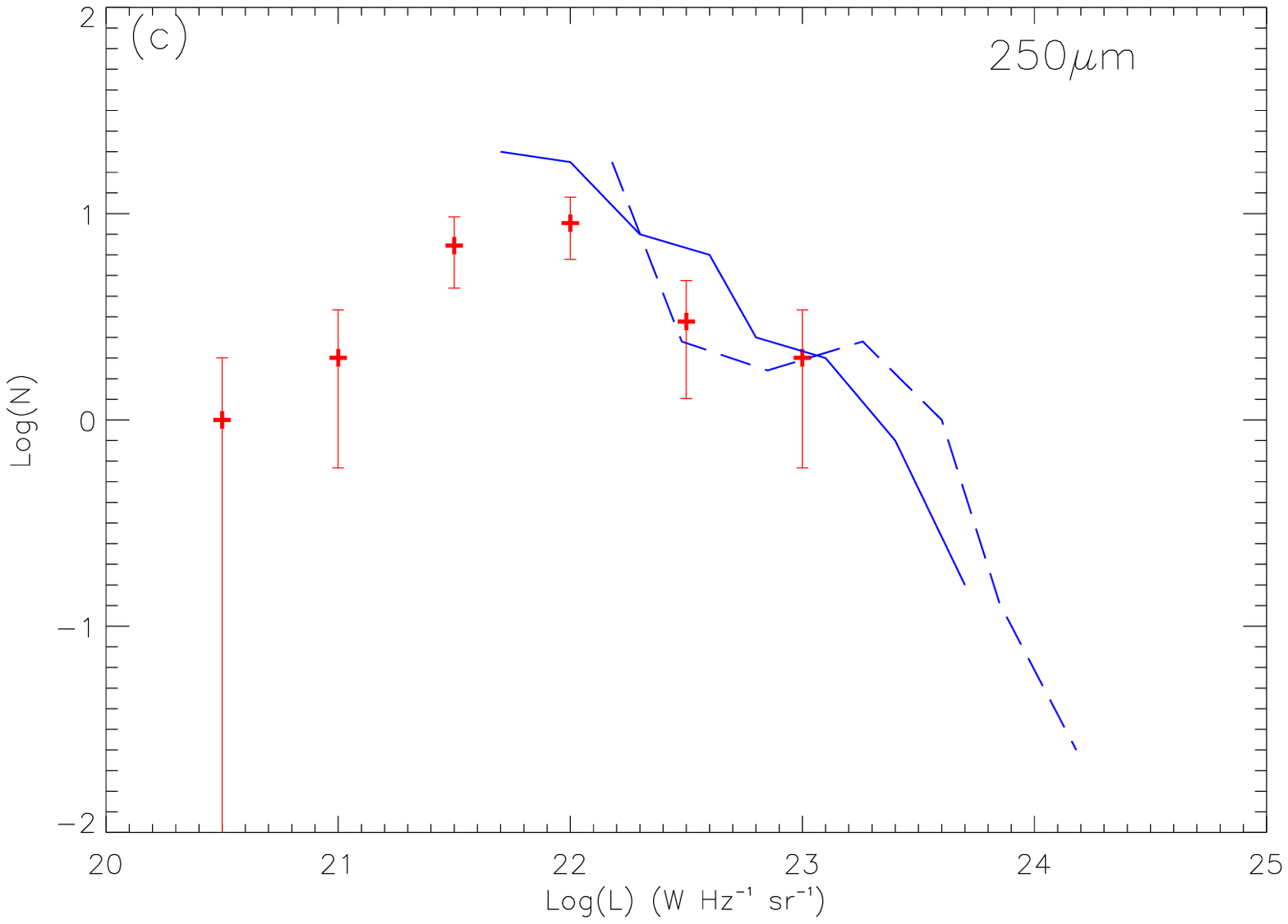}
\includegraphics[scale=0.38]{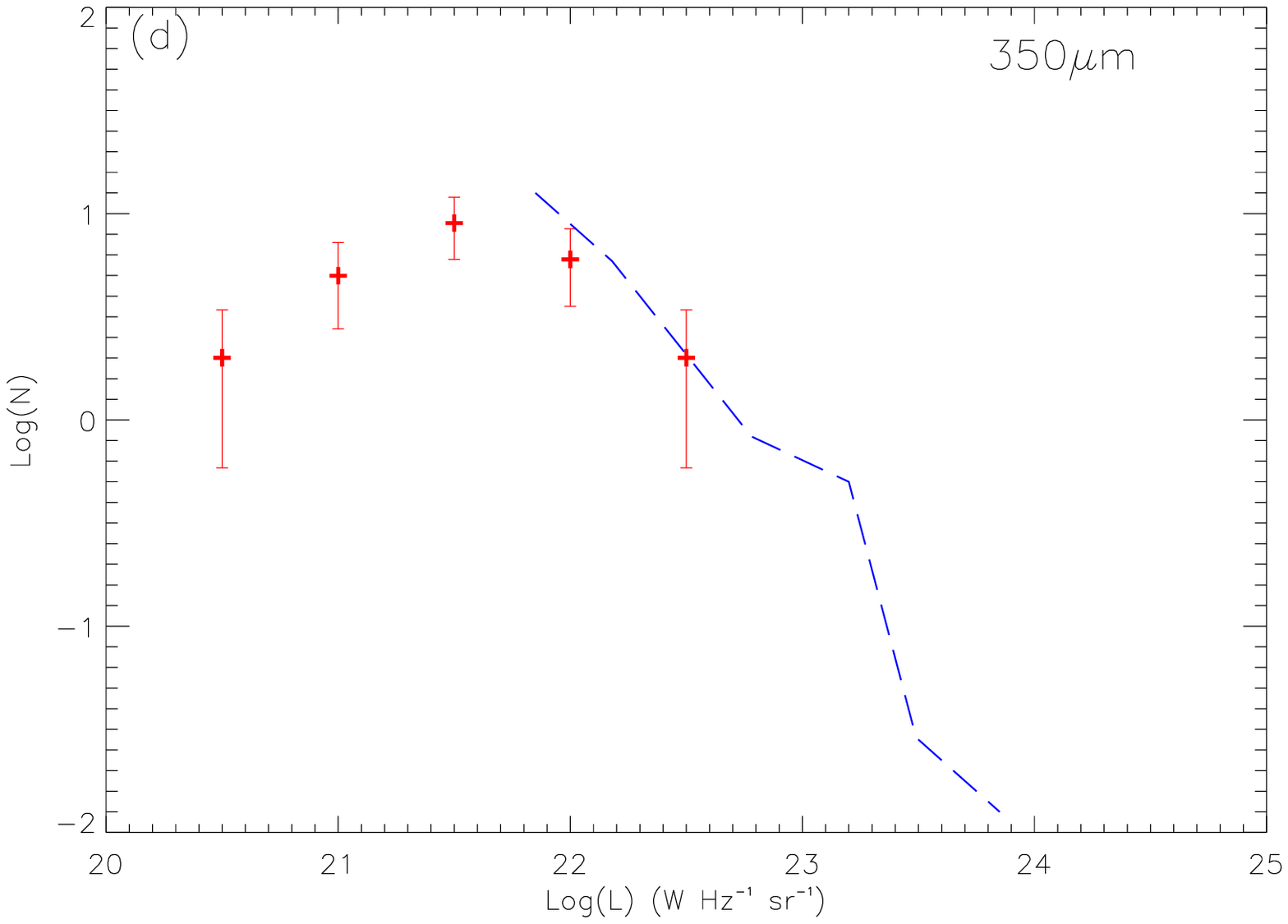}
\includegraphics[scale=0.38]{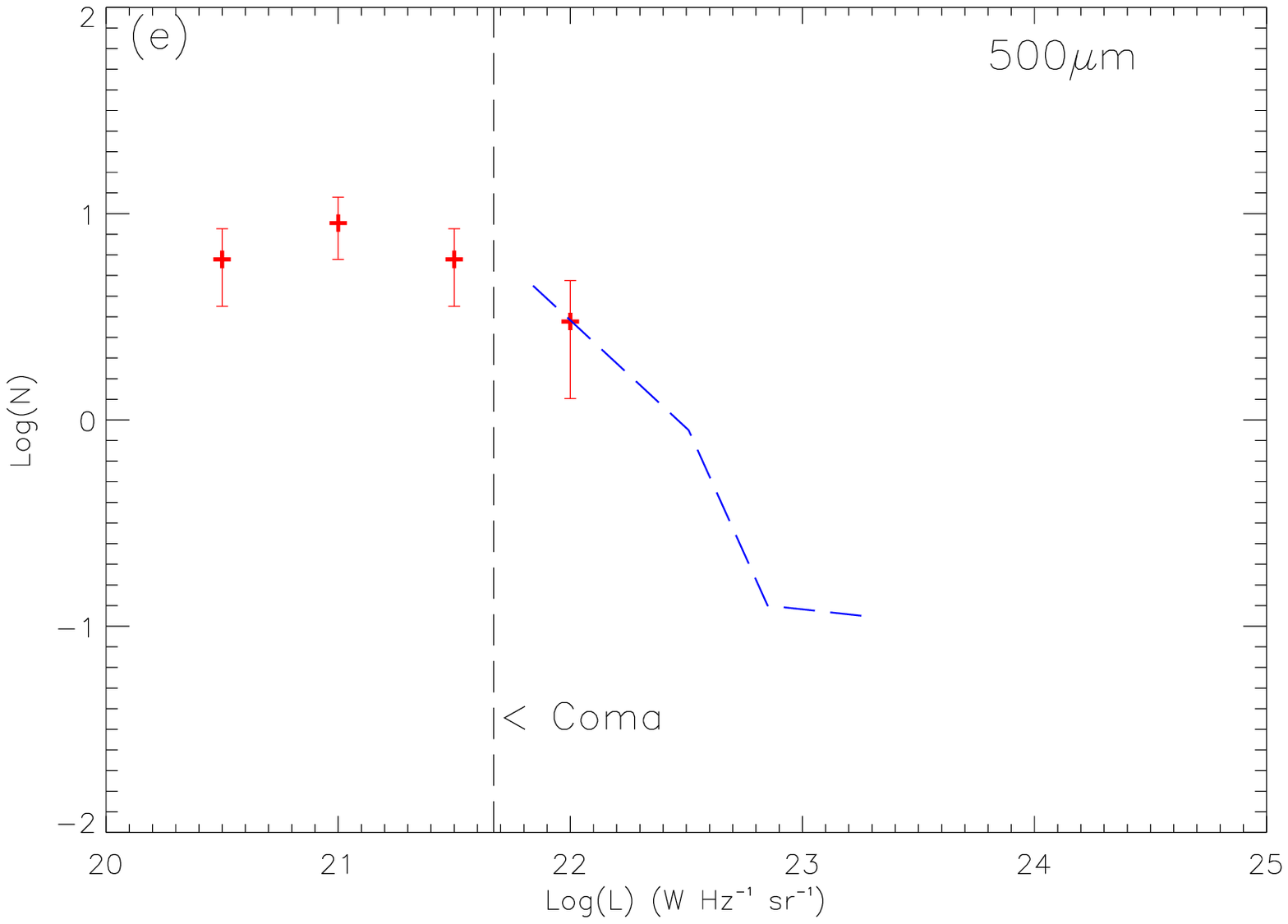}
\caption{
Luminosity functions at 100 to 500 $\mu$m for the Virgo Cluster bright galaxy sample (red crosses) with $\sqrt(N)$ counting errors. The various other lines and points are described in the text.
}
\end{figure}

For comparison we include some previously determined LFs (Fig. 3). Rowan-Robinson et al. (1987) derived an IRAS 100 $\mu$m LF for optically identified galaxies (over a wide range of environments) with photographic magnitudes less than 14.5 and an available redshift (blue dashed line Fig. 3a). This is roughly comparable with our Virgo sample in the sense that the galaxies all have to have an optical identification and a distance. The Rowan-Robinson et al., LF cuts off quite sharply at the bright end at about $10^{24}$ W Hz$^{-1}$ sr$^{-1}$, but then through to about $4\times10^{21}$ W Hz$^{-1}$ sr$^{-1}$ it follows a power law slope of $\approx$-1. Although the LF normalisation is arbitrary it is clear that the Rowan-Robinson et al., LF predicts $\sim10^{3}$ more galaxies at the faint end than at the bright end - even with the small numbers this is inconsistent with the Virgo data.  We suggest that this is most likely due to processes in the cluster that remove gas and dust, i.e. tidal and ram pressure stripping (Doyon \& Joseph, 1989). These processes are more effective for lower mass objects. The Rowan-Robinson et al., data also include galaxies that are one order of magnitude brighter than the brightest galaxies found in the cluster. Although in this sample there are very low numbers it has previously been noted that clusters in general do not have the very bright FIR sources seen in the field (Bicay \& Giovanelli, 1987). Also included in Fig. 3a (light blue stars) are the IRAS 100 $\mu$m data, i.e. the LF for all VCC galaxies in the field with an IRAS 100 $\mu$m detection (31 galaxies). Although subject to very different selection criteria the two data sets are broadly consistent with each other.

In Fig. 3b we show the derived LF at 160 $\mu$m. We have previously chosen to select our sample in a way that guarantees a detection in each of the five bands, but we are aware that observational selection may affect any inferences made. To investigate this we used the programme SExtractor to detect objects down to a diameter of 30\arcsec at a 1.5$\sigma_{160}$ isophote on the unsmoothed 160 $\mu$m data. Note that much of the Galactic cirrus is removed in the PACS data by the standard pipeline $1/f$ noise removal. The result is 112 detections of which 26 correspond with known VCC galaxies and are thus confirmed cluster members. These 26 are plotted as blue stars on Fig 3b, they are consistent with our previous sample. The complete (112) sample (assuming they all lie at the distance of Virgo) are also plotted as green triangles, the dashed line indicating a signal-to-noise of 5. The numbers rise steeply, but with a slope less than expected for the background counts (slope of -1.5). The nature of these galaxies and similar ones detected in other bands will be investigated in subsequent papers.

Eales et al., (2009) have used data from BLAST to derive 250, 350 and 500 $\mu$m LFs for galaxies at different redshifts. On Figs 3 c, d, e we compare the lowest redshift ($0<z<0.2$) LFs (blue dashed line) with the Virgo data (arbitray normalisation). There are some very luminous galaxies (for example $L_{250}>10^{24}$ W Hz$^{-1}$ sr$^{-1}$) in the BLAST data that are not found in Virgo, though their space densities are low at  $\sim 10^{-5}$ Mpc$^{-3}$. The BLAST LFs do not go beyond the peaks in the Virgo Cluster LFs and so it is difficult to say what the predicted numbers are at lower luminosities, but Eales et al. also show that the BLAST LFs are consistent with an extrapolation of the IRAS and SCUBA LFs, which do have a rising power law slope at the faint end. So our expectation is that we will again find that there is a lack of faint sources in Virgo when compared to galaxies in the general field. The solid blue line on Fig 3c is the recent Herschel-ATLAS (Eales et al., 2010) determination of the local 250 $\mu$m LF (Dye et al., 2010) which is roughly consistent with the BLAST data. We look forward to a future comparison with other galaxy clusters at these wavelengths.

Herschel large area surveys like ATLAS will eventually detect many more galaxy clusters. For example in the ATLAS fields there is the prominent and well studied Coma Cluster and $\sim$20 other Abell clusters with z$\le$0.1. The importance of the above Virgo data is that we can be quite sure that all of the galaxies are cluster members - this is not so for surveys of more distant clusters. Even so, with about 20 ATLAS clusters one would expect that the cluster LF will eventually be well measured, even though the greater distances may inhibit this analysis.
At the distance of the Coma Cluster, one of the nearest clusters in the ATLAS fields, (95 Mpc for H$_{0}$=73 km s$^{-1}$ Mpc$^{-1}$) our faintest 500 $\mu$m source (0.2 Jy at 17 Mpc) would have a flux of $\sim$6 mJy. This is below the ATLAS 500 $\mu$m flux limit of 53 mJy. The brightest source detectable by ATLAS at the distance of Coma would have a luminosity of $5\times10^{21}$ W Hz$^{-1}$ sr$^{-1}$, as indicated by the dashed line on Fig. 3(e). Thus almost 80\% of the Virgo Cluster galaxies in this sample would not be detected at the distance of Coma. The brightest source in Virgo with a 500 $\mu$m flux of $\sim$12 Jy would fail to be detected beyond a distance of 255 Mpc or z=0.06. Most of the clusters in the ATLAS fields lie within z=0.1, but at or beyond the distance of Coma, so our Virgo Cluster data predicts that only the very brightest cluster galaxies will be detected in the ATLAS. 

Using a single temperature modified blackbody ($\beta=2.0$) and an emissivity of 0.19 m$^{2}$ kg$^{-1}$ at 350 $\mu$m (Draine, 2003) we have fitted the SEDs of the 22 late-type galaxies in the sample. Dust masses and temperatures within these 500 $\mu$m determined apertures are in the range of $10^{5.9-7.7}$ M$_{\odot}$ and 15-24K respectively. This illustrates the potential sensitivity of the full depth survey to low dust masses ($<10^{5}$ M$_{\odot}$) and the existence of a significant cold dust component (T$<20$K). 

\section{Conclusions}
We have described the HeViCS and our first science data from PACS and SPIRE. The data were used to produce 100, 160, 250, 350 and 500 $\mu$m LFs for a sample of optically bright galaxies in the central region of the Virgo Cluster. Unlike the global optical LF and those previously derived in the FIR these functions do not appear to have a steep power law dependence at the faint end. There are relatively few faint FIR sources that can be associated with confirmed Virgo Cluster members compared to what is found for field galaxies and there are no examples of very luminous FIR sources.   \\


\begin{thebibliography}{}

\bibitem[Baes et al.(2010)]{Baes} Baes M., Clemens M., Xilouris E. M.,
  et al., A\&A, this volume
\bibitem[Bertin \& Arnouts(1996)]{Bertin} Bertin E., Arnouts S., 1996,
  A\&AS, 117, 393
\bibitem[Bicay et al.(1987)]{Bicay} Bicay M., Giovanelli R., 1987,
  ApJ, 321, 645
\bibitem[Binggeli et al.(1985)]{Binggeli} Binggeli B., et al., 1985,
  AJ, 90, 1681
\bibitem[Binggeli et al.(1987)]{Binggeli2} Binggeli B., Tammann G.,
  Sandage A., 1987, AJ, 94, 251
\bibitem[Binggeli et al.(1993)]{Binggeli3} Binggeli B., Popescu C.,
  Tammann G., 1993, A\&AS, 98, 275
\bibitem[Boehringer et al.(1994)]{Boeh} Boehringer H., et al., 1994,
  Nature, 368, 828
\bibitem[Boselli et al.(2010)]{Boselli} Boselli A., et al., A\&A,
  2010, this volume
\bibitem[Boselli \& Gavazzi(2006)]{BosGav} Boselli A., Gavazzi G.,
  2006, PASP, 118, 517
\bibitem[Clemens et al.(2010)]{Clemens} Clemens M., et al., 2010,
  A\&A, this volume
\bibitem[Cortese et al.(2010)]{Cortese} Cortese L., et al., 2010,
  A\&A, this volume
\bibitem[Davies et al.(2010)]{Davies} Davies J., et al., 2010, A\&A,
  this volume
\bibitem[Davies et al.(2004)]{Davies04} Davies J., et al., 2004,
  MNRAS, 349, 922
\bibitem[De Looze et al.(2010)]{De Looze} De Looze I., et al., 2010,
  A\&A, this volume
\bibitem[Draine(2003)]{Draine} Draine B., 2003 ARAA, 41, 241
\bibitem[Doyon \& Joseph(1989)]{Doyon} Doyon R., Joseph R., MNRAS,
  1989, 239, 347
\bibitem[Dye et al.(2010)]{Dye} Dye S., et al., 2010, A\&A, this
  volume
\bibitem[Eales et al.(2009)]{Eales} Eales S., et al., 2009, ApJ, 707,
  1779
\bibitem[Eales et al.(2010)]{Eales2010} Eales S., et al., 2010, PASP,
  in press
\bibitem[Gavazzi et al.(1999)]{Gav99} Gavazzi G., Boselli A.,
  Scodeggio M., Pierini D. and Belsole E., 1999, MNRAS, 304, 595
\bibitem[Gavazzi et al.(2003)]{Gav03} Gavazzi, G. Boselli, A. Donati,
  A. Franzetti, P. and Scodeggio, M., 2003, A\&A, 400, 451
\bibitem[Girardi et al.(1998)]{Gir98} Girardi M., Giuricin G.,
  Mardirossian F., Mezzetti M. and Boschin W., 1998, ApJ, 505, 74
\bibitem[Gomez et al.(2003)]{Gomez} Gomez et al., 2003, ApJ, 584, 210
\bibitem[Griffin et al.(2010)]{Griffin} Griffin M., et al., 2010, A\&A,
  this volume
\bibitem[Grossi et al.(2010)]{Grossi} Grossi M., et al., 2010, A\&A,
  this volume
\bibitem[Haynes \& Gionvanelli(1984)]{Haynes} Haynes M., Giovanelli
  R., 1984, AJ, 89, 758
\bibitem[Lewis et al.(2002)]{Lewis} Lewis I., et al., 2002, MNRAS,
  334, 673
\bibitem[Pascale et al.(2010)]{Pas} Pascale E., et al., 2010, A\&A,
  this volume
\bibitem[Pilbratt et al.(2010)]{Pil} Pilbratt G., et al., 2010, A\&A,
  this volume
\bibitem[Poglitsch et al.(2010)]{Pog} Poglitsch A., et al., 2010,
  A\&A, this volume
\bibitem[Pohlen et al.(2010)]{Poh} Pohlen M., et al., 2010, A\&A, this
  volume
\bibitem[Rines \& Diaferio(2006)]{Rin} Rines K., Diaferio A., 2006,
  AJ, 132, 1275
\bibitem[Rowan-Robinson et al.(1987)]{Row} Rowan-Robinson M., Helou
  G., Walker D., 1987, MNRAS, 227, 589
\bibitem[Skillman et al.(1996)]{Ski} Skillman E., Kennicutt R.,
  Shields G., Zaritsky D., 1996, ApJ, 462
\bibitem[Smith et al.(2010)]{Smi} Smith M., et al., 2010, A\&A, this
  volume
\bibitem[Swinyard et al.(2010)]{Swi} Swinyard B., et al., 2010, A\&A,
  this volume

\end{thebibliography}
\end{document}